\documentclass[review,3p]{elsarticle}

\usepackage{hyperref}

\usepackage{libertinus}
\usepackage{libertinust1math}
\usepackage{courier}
\usepackage[T1]{fontenc}
\DeclareUrlCommand\path{\urlstyle{sf}}
\makeatletter
\let\@afterindenttrue\@afterindentfalse
\makeatother
\usepackage{booktabs,rotating,multicol,multirow,textgreek}
\DeclareUnicodeCharacter{03BA}{\textkappa}
\usepackage{xcolor}

\nopreprintlinetrue

\begin{document}

\begin{frontmatter}

\title{Evaluating Generic Auto-ML Tools for Computational Pathology}

%% or include affiliations in footnotes:
\author[mevisaddress]{Lars Ole Schwen\corref{correspondingauthor}}
\cortext[correspondingauthor]{Corresponding author}
\ead{ole.schwen@mevis.fraunhofer.de}

\author[mevisaddress]{Daniela Schacherer}
\author[daiaddress]{Christian Geißler}
\author[mevisaddress]{André Homeyer}

\address[mevisaddress]{Fraunhofer Institute for Digital Medicine MEVIS, Max-von-Laue-Str. 2, 28359 Bremen, Germany}
\address[daiaddress]{DAI‑Labor, Technische Universität Berlin, Ernst‑Reuter‑Platz 7, 10587 Berlin, Germany}

\begin{abstract}
  Image analysis tasks in computational pathology are commonly solved
  using convolutional neural networks (CNNs). The selection of a
  suitable CNN architecture and hyperparameters is usually done
  through exploratory iterative optimization, which is computationally
  expensive and requires substantial manual work. The goal of this
  article is to evaluate how generic tools for neural network
  architecture search and hyperparameter optimization perform for
  common use cases in computational pathology.  For this purpose, we
  evaluated one on-premises and one cloud-based tool for three
  different classification tasks for histological images: tissue
  classification, mutation prediction, and grading.

  We found that the default CNN architectures and parameterizations of
  the evaluated AutoML tools already yielded classification
  performance on par with the original publications. Hyperparameter
  optimization for these tasks did not substantially improve
  performance, despite the additional computational effort. However,
  performance varied substantially between classifiers obtained from
  individual AutoML runs due to non-deterministic effects.

  Generic CNN architectures and AutoML tools could thus be a viable
  alternative to manually optimizing CNN architectures and
  parametrizations. This would allow developers of software solutions
  for computational pathology to focus efforts on harder-to-automate
  tasks such as data curation.
\end{abstract}

\begin{keyword}
  computational pathology \sep
  convolutional neural networks \sep
  AutoML \sep
  hyperparameter optimization \sep
  reproducibility
\end{keyword}

\end{frontmatter}

\section{Introduction}

\subsection{Motivation}

Automated analysis of digital tissue images holds great potential for
improving pathology diagnostics. In addition to increasing the
efficiency and reproducibility of disease assessment, it also enables
the extraction of novel digital biomarkers for predicting outcomes and
treatment response~\citep{AcsRanHar2020, BerSchRim2019}.  Image
analysis tasks in computational pathology are typically solved using
Convolutional Neural Networks (CNNs)~\citep{Par2019, CuiZha2021,
  CamHanGen2019, KomIsh2019, JanMad2016, AbePanAef2019}.  Developing
such CNN-based solutions~\citep{JanMad2016} requires multiple steps
including data curation and the translation between the diagnostic
task and the bare machine learning task.

An essential task in developing a CNN-based solution is the choice of
a suitable neural network architecture, its initialization, and
(training) hyperparameters~\citep{RitWolBer2019, BerKomEli2015,
  LiaLiaNis2018, RenXiaCha2021}.  CNN architectures and
hyperparameters are usually selected based on experience and manual
optimization, where different architectures are tried, multiple
trainings are performed, and hyperparameters are adjusted. This
process is very time consuming and computationally intensive for a
number of reasons: The optimization task is multi-dimensional,
involving categorical (e.g., CNN architecture), Boolean (e.g., whether
to use certain CNN building blocks), integer (e.g., number of epochs
or early stopping patience), and continuous (e.g., learning rate)
variables. Evaluating the objective function involves computing a
metric after training and evaluating a CNN. This is computationally
very expensive, single evaluations may take hours to days to
compute. The objective function is neither convex nor is a gradient
(approximation) available. Moreover, the objective function is noisy
due to non-determinism in CNN training~\citep{PhaQiaWan2020,
  MorWil2020, NagWarSto2018}.

\subsection{State of the art}

For some time, research has been carried out on methods to automate
the search for optimal CNN architectures and hyperparameters and make
these steps more efficient. Such “AutoML” methods have been
published for generic applications~\citep{ElsSak2019, ElsMahSak2019}
and medical image analysis~\citep{RitWolBer2019}.  Their main building
blocks are automatic neural architecture search~\citep{PhaGuaZop2018,
  ElsMetHut2019, ZopVasShl2018} and algorithmic hyperparameter
optimization~\citep{BerKomEli2015, BerKomEli2015}.

A number of cloud-based generic commercial AutoML tools is available,
e.g., Amazon Sage Maker~\citep{web-AWSImageClassification},
Clarifai~\citep{web-clarifai}, Google AutoML
Vision~\citep{web-googleautoml, Bis2019} and Vertex
AI~\citep{web-vertexai}, H2.ai~\citep{web-h2oai}
MedicMind~\citep{web-medicmind}, as well as Microsoft Azure Custom
Vision~\citep{web-customvision}.  Besides offering cloud deployment for
inference, limited possibilities to download trained classifiers for
on-premises use exist as well.  Moreover, there is Apple Create
ML~\citep{web-applecreateml} for offline application of AutoML. These
tools are designed as easy-to-use tools for machine learning tasks, in
particular image classification, without substantial user
configuration~\citep{ObyAbbKor2021} or the need for coding
experience~\citep{FaeWagFux2019}.  Instead, usage requires providing
image data and tables with reference results, CNN training is
subsequently started via graphical (web) interfaces. While mostly
evaluated on standard benchmark tasks like ImageNet or CIFAR-10/-100
(e.g.,~\citep{HexZhaChu2021}), cloud-based generic
AutoML tools have also been evaluated in medical image
analysis~\citep{BorVisTho2020, KorGuaFer2021, KimLeePar2021,
  GhoTanChu2021, WanWonIpx2021, SakFerKal2020} and specifically in
computational pathology~\citep{ZenZha2020, BorWilBor2019a, Pur2020}.

There are also Auto ML tools for pure on-premises use, such as
AutoGluon~\citep{web-autogluon, EriMueShi2020, web-MediumAutoGluon},
AutoKeras~\citep{web-autokeras, JinSonHux2019, Yex2020},
Auto-PyTorch~\citep{web-autopytorch, ZimLinHut2021}, and
model\_search~\citep{web-modelsearch}.  These open-source tools are
used as part of code (e.g., Python) and are typically more flexible,
requiring more user configuration than the cloud-based tools. Besides
generic tools, specialized CNN training tools for biological
microscopy image segmentation~\citep{BelJok2021} and histopathology
image analysis~\citep{WanCouZha2021} have been published.

\subsection{Contribution}

In this article, we evaluate how generic AutoML tools for image
classification can be applied to selected use cases in computational
pathology. Moreover, we investigate how the resulting classifiers
perform compared to reference results from literature for the selected
tasks.

\section{Methods}

\subsection{AutoML tools}

We evaluated the performance of two generic AutoML tools (AutoGluon
version 0.1.0~\citep{web-autogluon-pypi} and AutoML
Vision~\citep{web-googleautoml} as available between Sep~02 and Sep~22,
2021) for image classification when applied to three different use
cases~\citep{CouOcaSak2018, KatPeaHal2019, ArvFriMor2018} in
computational pathology. By “performance,” we refer to the
“correctness” of the classification results evaluated in
task-specific metrics. The evaluated tools were chosen to include one
cloud-based and one on-premises tool, both allowing on-premises
inference using the trained CNNs. The latter is important if the
classifier shall be applied to confidential data, to avoid vendor
lock-in by cloud providers, and to allow for a defined hardware plus
software setup to ensure reliability, e.g., as required for regulatory
approval if the CNN becomes part of a medical product. With AutoGluon
originally developed by Amazon~\citep{EriMueShi2020, web-awsautogluon}
and AutoML Vision developed by Google, these tools come from two
different major platform providers.

\subsection{Example use cases}

To evaluate these AutoML tools, we chose the following three use cases
covering different types of diagnostic tasks in computational
pathology.  All three are based on frequently cited publications, so
that reference values for the performance assessment were
available. Moreover, we required that the underlying datasets were
publicly available. All three use cases translate to image
classification, the most common application of machine learning in
computational pathology.

\subsubsection{Tissue classification: lung cancer}

From~\citep{CouOcaSak2018}, we selected two binary and one ternary
classification task, distinguishing normal vs.\ tumor tissue, lung
adenocarcinoma (LUAD) vs.\ lung squamous cell carcinoma (LUSC) tissue,
and normal vs.\ LUAD vs.\ LUSC tissue. The authors split data obtained
from The Cancer Genome Atlas (TCGA)~\citep{web-tcga} in training,
validation, and test data at the slide level, divided the whole-slide
images in image tiles, and trained separate Inceptionv3
CNNs~\citep{SzeVanIof2016} for each image classification task. In these
datasets, one class was assigned to all tiles of the entire slide,
even if the slide also contained some parts of healthy tissue. The
authors evaluated classifier performance at the slide level, averaging
prediction probabilities over all tiles belonging to a slide, via the
area under the receiver operating curve (AUROC), resulting in 0.993
(for the binary classification normal vs.\ tumor), 0.950 (LUAD
vs.\ LUSC), 0.984/0.969/0.966 (three-class normal vs.\ rest/LUAD
vs.\ rest/LUSC vs.\  rest). The code for~\citep{CouOcaSak2018} is
available from~\citep{web-CoudrayCode}.

\subsubsection{Mutation prediction: microsatellite instability}

From~\citep{KatPeaHal2019}, we selected the binary classification task
of predicting microsatellite instability (MSI) vs.\ microsatellite
stability (MSS) from tissue samples. The authors applied this to three
different dataset: tiled whole-slide images of formalin-fixed
paraffin-embedded tissue samples of colorectal cancer (CRC-DX),
snap-frozen samples of colorectal cancer (CRC-KR), and formalin-fixed
paraffin-embedded tissue samples of gastric (stomach) adenocarcinoma
(STAD). The data obtained from TCGA was first split at the patient
level in two datasets, one for training and validation to be split
further, and one for testing. The tile images split in this way are
available from~\citep{Kat2019b} (CRD-DX, STAD) and~\citep{Kat2019a}
(CRC-KR). Next, training and validation data were split at the tile
level 85\%:12.5\%, the remaining 2.5\% of the tiles remained
unused~\citep{web-KatherCode}, in order to train separate ResNet18
CNNs~\citep{HexZhaRen2016} for classifying each dataset. Like in the
tissue classification task~\citep{CouOcaSak2018}, the authors
of~\citep{KatPeaHal2019} evaluated the classifier performance per slide
(after averaging prediction probabilities over all tiles for a given
slide) in terms of AUROC, resulting in 0.84 (for the CRC-DX dataset),
0.77 (CRC-KR), and 0.81 (STAD).

\subsubsection{Grading: Gleason grading from tissue microarrays}

From~\citep{ArvFriMor2018}, we selected the quaternary classification
task of distinguishing Gleason grades 0, 3, 4, and 5 from tissue
microarray (TMA) images of prostate biopsies. The authors split the
dataset~\citep{ArvFriMor2018a} into training, validation, and test data
at the TMA level. Regions with different Gleason grades were annotated
in each TMA spot by one pathologist; an additional pathologist
annotated the test data.

This evaluation differs from the other two use cases: One TMA spot may
have regions with different Gleason grades, so only an evaluation at
the image tile level makes sense. For this purpose, the authors
divided the TMA spots in image tiles of 750² pixels and either
assigned the single grade annotated for the center 250² pixels or
discarded the tile in case of non-unique annotation for the
center. Moreover, the authors used quadratically weighted Cohen's~κ as
a classifier performance metric, reflecting that consecutive grades
permit a more fine-grained evaluation than categorical classes.

The authors trained a MobileNet CNN~\citep{HowZhuChe2017} and obtained
κ values of 0.55 (compared to pathologist~1 who also annotated the
training and validation data) and 0.49 (compared to
pathologist~2). Note that these values cannot be compared
quantitatively to the AUROC values for the other two use cases, but
can be compared to the inter-pathologist agreement of
0.66. Subsequently, the authors of~\citep{ArvFriMor2018} used a
sliding-window approach to obtain a pixel-wise classification,
computed a Gleason score per TMA spot and obtained κ values of 0.75
(algorithm vs.\ pathologist~1) and 0.72 (algorithm vs.\ pathologist~2),
comparable to an inter-observer agreement of 0.75 (pathologist~1
vs.\ 2).  However, we decided to omit this aggregation step in the
evaluation here in order to have one use-case focused on the plain
tile classification.

For the present study, this use case also illustrates how data
augmentation~\citep{TelLitBan2019b} can be implemented as a
preprocessing step before using the evaluated AutoML tools.

\subsection{Data preparation}

\subsubsection{Tissue classification}

For the tissue classification use case, we obtained whole-slide images
(WSI) at 20× objective magnification from the Genomic Data
Commons~\citep{web-GDC} as specified in the corresponding download
manifest~\citep{web-CoudrayCode}.  We did not encounter readability
issues as mentioned in~\citep{CouOcaSak2018} and used all 2167
downloaded slides. A corresponding metadata file was also available
which specified the tissue type (normal, LUAD, LUSC) for each WSI. For
each slide, we generated 512²\,px tiles, omitting tiles that show
\textgreater{} 50\% background as described
in~\citep{web-CoudrayCode}: A pixel is considered background if its
8-bit RGB values satisfy $0.299R+0.587G+0.114B > 220$. For reducing
computational effort, we subsequently downsampled the tiles to
128²\,px using PIL Image.resize()~\citep{web-pillow}, saving them as
JPEG images at quality 75. Preliminary empirical experiments indicated
that this way of using the same field of view with fewer image details
led to comparable classification quality. The dataset was split at the
patient level with the target of training, validation, and test set
comprising 70\%, 15\%, and 15\% of the tiles, respectively. We ensured
that all tiles from a particular patient are assigned to only one
set. This split was thus at a higher level than the slide-level split
in~\citep{CouOcaSak2018}, making the task slightly harder. For a
detailed list of the numbers of patients, slides, and image tiles in
this dataset, we refer to
\autoref{suppTab:tileStatisticsTissueClassification} in the
appendix. This tile preparation and dataset split were run once as a
preprocessing step.

\subsubsection{Mutation prediction}

For the mutation prediction use case, we downloaded the datasets
from~\citep{Kat2019b, Kat2019a}.  The test sets are fixed as separate
archives, with file names indicating a split at the patient level. To
divide into training and validation data, we split the non-test data
at the tile level according to the percentages described
in~\citep{KatPeaHal2019}: 85\% training and 12.5\% validation, with
2.5\% of the tiles remaining unused. The second split was done
separately per AutoML run, so different trainings used different
splits in training, validation, and unused data. For the numbers of
patients, slides, and image tiles in this dataset, we refer to
\autoref{suppTab:tileStatisticsMutationPrediction} in the appendix.

\subsubsection{Grading}

For the grading use case, we downloaded the dataset
from~\citep{ArvFriMor2018a} and pre-processed the images as described
in~\citep{ArvFriMor2018}.  For this purpose, we technically adapted the
preprocessing script from~\citep{web-ArvanitiCode} and downsampled the
images using the ImageMagick~\citep{web-imagemagick} convert command
line tool. After resampling all images from 750² to 250² pixels,
validation and test tiles were cropped to the center 224² pixels.

We used this use case as an example for how data augmentation and
balancing needs to be implemented as a preprocessing step before using
the AutoML tools. For data augmentation, we cropped 224² pixel images
at random position in the 250² images, randomly applied or omitted
horizontal mirroring, vertical mirroring, and 90-degrees rotation (all
implemented as numpy array~\citep{web-numpy} operations), as well as
color augmentation in the hue–saturation–value (HSV) color space
(which is a realistic variation for histological images and
recommended by~\citep{TelLitBan2019b}; implemented using the
albumentations Python package~\citep{BusIglKhv2020,
  web-albumentations} with maximal shifts of 20, 30, and 20 for H, S,
and V, respectively). We combined 8-fold data augmentation with
oversampling the minority classes. For the numbers of image tiles in
this dataset, we refer to \autoref{suppTab:tileStatisticsGrading} in
the appendix. For implementational convenience, these steps of tile
preparation were run as a preprocessing prior to each AutoGluon run
and not just once. By seeding the pseudo-random number generator used
for cropping and augmentation, however, we ensured that all trainings
were run on the same tile data.

\subsubsection{Use with AutoGluon}

Datasets in AutoGluon were loaded by specifying a folder with
subfolders for each class containing the image tiles. Loading all
image tiles at once turned out to be impossible with 32 GiB of
RAM. Hence, we randomly split each full dataset in 40 subsets
(folders) during preparation, loaded these subsets successively as
datasets for use with AutoGluon, and concatenated them to a single
dataset.

\subsubsection{Use with AutoML Vision}

For using datasets with AutoML Vision, image tiles needed to be
available in a Google Storage Bucket with a list describing the split
in training, validation, and test data as well as the reference
classes (in csv, comma-separated values, format). We hence created
these csv lists and uploaded them together with the individual image
tiles.

The size of a dataset for each cloud AutoML task was limited to one
million images. For the mutation prediction and grading use cases,
this limit was not reached; we used the same training, validation, and
test sets as for AutoGluon. Adaptations were necessary for the tissue
classification use case: training and validation data for the “LUAD
vs.\  LUSC” task contained less than one million tiles, so we only
reduced the number of test images by random sampling from the original
test set.  This does not change the task, as we do not use the
tile-based performance reported by the cloud tool, but evaluate the
actual performance metric on premises as a postprocessing step using
the full test set. The “normal vs.\ tumor” and “normal vs.\ LUAD
vs.\ LUSC” tasks had more than one million training and validation
tiles, so we randomly sampled 820850, 179050, and 100 tiles from the
original training, validation, and test sets, respectively. These
numbers were chosen such that the relevant sets, training and
validation, were kept large and size proportionality between them was
preserved. This way, the task became harder, as less training data was
used.

\subsection{Generating convolutional neural networks}

\subsubsection{On-premises training using AutoGluon}

Based on the available presets for AutoGluon, we evaluated different
presets for different CNN architectures with limited time budget for
training (up to 6 hours wall-clock time), in order to get a first
overview of the performance of the resulting classifiers for all use
cases. We then selected the generally most promising preset for a
limited hyperparameter optimization (with a time budget for training
of 14 days) for one task per use case (LUAD vs.\ LUSC for the tissue
classification, the STAD dataset for the mutation prediction, and the
single Gleason grading task) in order to limit the computational
workload. The presets used in this evaluation are listed in
\autoref{tab:AutoGluonPresets}, where the budgets refer to the
actual training and exclude preparation steps (e.g., loading datasets)
and postprocessing steps (e.g., applying the classifier to the test
set). We stored the console output from the training for later
analysis and saved the resulting CNN for later evaluation.

\begin{table}
  \centering
  \caption{Presets used for the evaluation of AutoGluon (ESP: Early
    stopping patience, HPO: algorithm for hyperparameter optimization,
    Budget: training time limit in wall-clock time in hours, BayesOpt:
    Bayesian optimization as implemented in AutoGluon)}
    \label{tab:AutoGluonPresets}

  \footnotesize
  \begin{tabular}{llllllll}
    \toprule
    Name & Architecture       & Learning rate                              & Batch size             & \# Epochs   & ESP & HPO                 & Budget        \\
    \midrule
    0    & ResNet50\_v1b      & 0.01                                       & 64                     & 50          & 5   & none                & 1             \\
    1    & MobileNetv3\_small & \{0.01, 0.005, 0.001\}                     & \{64, 128\}            & \{50, 100\} & 10  & BayesOpt, 12 trials & 6             \\
    2    & ResNet18\_v1b      & \{0.01, 0.005, 0.001\}                     & \{64, 128\}            & \{50, 100\} & 10  & BayesOpt, 12 trials & 6             \\
    3    & ResNet34\_v1b      & [0.0001, 0.01]\textsubscript{logarithmic}  & \{8, 16, 32, 64, 128\} & 150         & 20  & BayesOpt, 16 trials & 6             \\
    4    & ResNet50\_v1b      & 0.01                                       & 64                     & 50          & 5   & none                & 6             \\
    5    & ResNet101\_v1d     & [0.00001, 0.01]\textsubscript{logarithmic} & \{32, 64, 128\}        & 200         & 50  & BayesOpt, 16 trials & 6             \\
    HPO  & ResNet18\_v1b      & [0.0001, 0.01]\textsubscript{logarithmic}  & \{64, 128\}            & \{50, 100\} & 10  & BayesOpt, 16 trials & 336 (14 days) \\
    \bottomrule
  \end{tabular}
\end{table}

\subsubsection{Cloud-based training using AutoML Vision}

AutoML Vision provided three presets for training CNNs which can be
downloaded for offline use. Depending on the dataset size, a training
time budget was recommended for the chosen preset. In addition to the
recommended budgets for the three presets, we also used the default
preset with a training budget limited to 2 node-hours. These presets
are listed in \autoref{tab:AutoMLVisionPresets}. Further
configuration of the training was neither needed nor possible.

\begin{table}
  \centering
  \caption{Presets used for the evaluation of AutoML Vision}
  \label{tab:AutoMLVisionPresets}

  \footnotesize
  \begin{tabular}{llll}
    \toprule
    Preset name (used here) & Name (during setup) & Name (after training) & Training time limit in node hours              \\
    \midrule
    A                       & Best Tradeoff       & Mobile Best Tradeoff  & 2                                              \\
    B                       & Best Tradeoff       & Mobile Best Tradeoff  & 14 (mutation prediction and grading use cases) \\
                            &                     &                       & 16 (tissue classification use case)            \\
    C                       & Faster Predictions  & Mobile Low Latency    & 14 (mutation prediction and grading use cases) \\
                            &                     &                       & 16 (tissue classification use case)            \\
    D                       & Higher Accuracy     & Mobile High Accuracy  & 16 (mutation prediction and grading use cases) \\
                            &                     &                       & 18 (tissue classification use case)            \\
    \bottomrule
  \end{tabular}
\end{table}

\subsection{Tile-based inference}

\subsubsection{On-premises inference using AutoGluon}

AutoGluon permitted loading the classifier trained in the previous
step and loading datasets for inference in the same way as for
training. The trained classifier was applied and the resulting
predicted classes/prediction probabilities for the individual image
file names were saved as csv lists. This worked in the same way for
all three use cases, using the previously prepared datasets. We ran
the inference separately for all training, validation, and test
datasets, storing the respective csv files for later evaluation.

\subsubsection{On-premises inference using AutoML Vision}

AutoML Vision permitted exporting the trained models to TensorFlow
Lite format (among other formats) for offline use. This resulted in an
integer-quantized CNN which could be downloaded from an associated
Google storage bucket. We loaded these CNNs in a Python program to run
the inference, this time on the test data only. With custom Python
code, we saved predictions and prediction probabilities in the same
csv format as produced by AutoGluon.

\subsection{Evaluation of classifier performance}

The tile-based results of the classifiers created with both AutoML
tools were used as the basis for calculating the specific performance
metrics for each use case.

The tissue classification and mutation prediction use cases used the
same evaluation metric, i.e., slide-based AUROC~\citep{CouOcaSak2018,
  KatPeaHal2019}.  This involved aggregating tile-based classification
results to a slide-based result by arithmetically averaging class-wise
prediction probabilities, where the tile-to-class mapping was
available from the TCGA file names. Then, we compute AUROCs from the
slide-based class prediction probabilities using the Python module
Scikit-learn~\citep{PedVarGra2011, web-scikitlearn}.  For the ternary
classification task, each class was separately compared to the
combination of the other two classes, i.e., we computed multiclass
one-vs-rest ROCs.

For the grading use case, quadratically weighted Cohen's~κ was used as
the performance metric. We computed these values using the Python
module Scikit-learn~\citep{PedVarGra2011, web-scikitlearn}, comparing
the output of the classifiers separately to the classifications of the
two pathologists who annotated the test data.

\subsection{Implementation and hardware}

For the evaluation of AutoGluon, the entire processing pipeline was
implemented as custom shell and Python scripts and executed in a
Docker~\citep{NusSocMar2020} container using one GPU (NVIDIA GeForce
GTX 1080 Ti or RTX 2080 Ti with 11178 or 11019 MiB GPU RAM).

For use with AutoML Vision, each of the preprocessed datasets was
exported from the preprocessing as part of the AutoGluon pipeline. The
on-premises inference using the downloaded classifiers from AutoML
Vision was implemented using Python code, the respective metrics were
evaluated using the same code as above. These scripts were executed in
a Docker container running locally on a PC with one NVIDIA GeForce GTX
1080 GPU with 8117 MiB GPU RAM.

\subsection{Evaluation of hyperparameter optimization}

For evaluating the effect of the hyperparameter optimization (HPO), we
analyzed the AutoGluon console outputs for the preset~HPO. AutoGluon
printed status information during its trials, but only saved what it
found as the best classifier at the end. Hence, classifier performance
could only be assessed on a tile level for the training and validation
data, but neither for the test data nor via slide-based performance
metrics. From the console output, we parsed the values of the
hyperparameters being optimized, the processing speed (images per
second; per batch and per epoch), the training accuracy (per batch and
per epoch), and the validation accuracy (per epoch and at the end of
each trial, after potential early stopping).

\subsection{Evaluation of reproducibility}

Since training and evaluation of CNN-based classifiers are not fully
deterministic processes~\citep{PhaQiaWan2020, MorWil2020,
  NagWarSto2018, LixTal2020, LinHut2020}, it is important to examine
whether differences in the performance of classifiers are meaningful
or merely random effects. For assessing the impact of these effects on
classifier performance, we repeated the entire AutoML run and
evaluation for selected presets six times in addition to the initial
run for the same tasks selected for hyperparameter optimization. In
these repetitions, we refrained from reducing variability for those
effects we could have controlled (e.g., seeding pseudo-random number
generators or avoiding concurrent system load) in order to have a
scenario better resembling the attempt to reproduce literature
results. Given the relatively small number of samples and, in
particular, just a single reference value per reproducibility
experiment, we omitted a potentially misleading assessment of
statistical significance.

\section{Results}

\subsection{Usage of the AutoML tools}

AutoGluon was relatively easy to set up for usage in Docker: a current
nvidia/cuda image needed only a few additional system and python
packages. AutoGluon starts the training of a CNN of the specified
architecture pre-trained on ImageNet~\citep{DenDonSoc2009} and saves
the trained CNN in a custom format of Gluon for later use. The presets
with budgets of 1h and 6h did not complete the planned number of
trials in the respective time budget, the hyperparameter optimization
presets with a budget of 14 days finished all 16 trials within 11d
12:17h for the tissue classification use case, 1d 17:11h for the
mutation prediction use case, and 1d 16:06h for the grading use case
(sic: tissue classification took ten days longer than the other two
use cases).

AutoML Vision did not provide any log of the initialization, potential
network architecture search, or hyperparameter optimization. The
platform allowed downloading the trained CNN in TensorFlow Lite format
(among other formats). The resulting networks appeared to be variants
of MnasNet~\citep{TanChePan2019}, with the same overall structure. The
number of channels (and thus number of network parameters) differed
throughout the building blocks of the CNN depending on the preset and,
in the final layers, depending on the number of classes. The network
outputs 8-bit (unsigned integer) quantized class probabilities. The
AutoML runs using presets~B to~D (see
\autoref{tab:AutoMLVisionPresets}) typically took less than half the
specified time budget, i.e., finished within a similar time as the
budget used for AutoGluon.

The dockerized code for preprocessing data, running AutoGluon to
generate CNNs (AutoGluon only), and evaluating the resulting
classifiers (AutoGluon and AutoML Vision) is available
from~\cite{codeSupplement}.

\subsection{Performance of classifiers obtained from presets}

\subsubsection{AutoGluon}

\begin{figure}[p]
  \includegraphics[width=\textwidth]{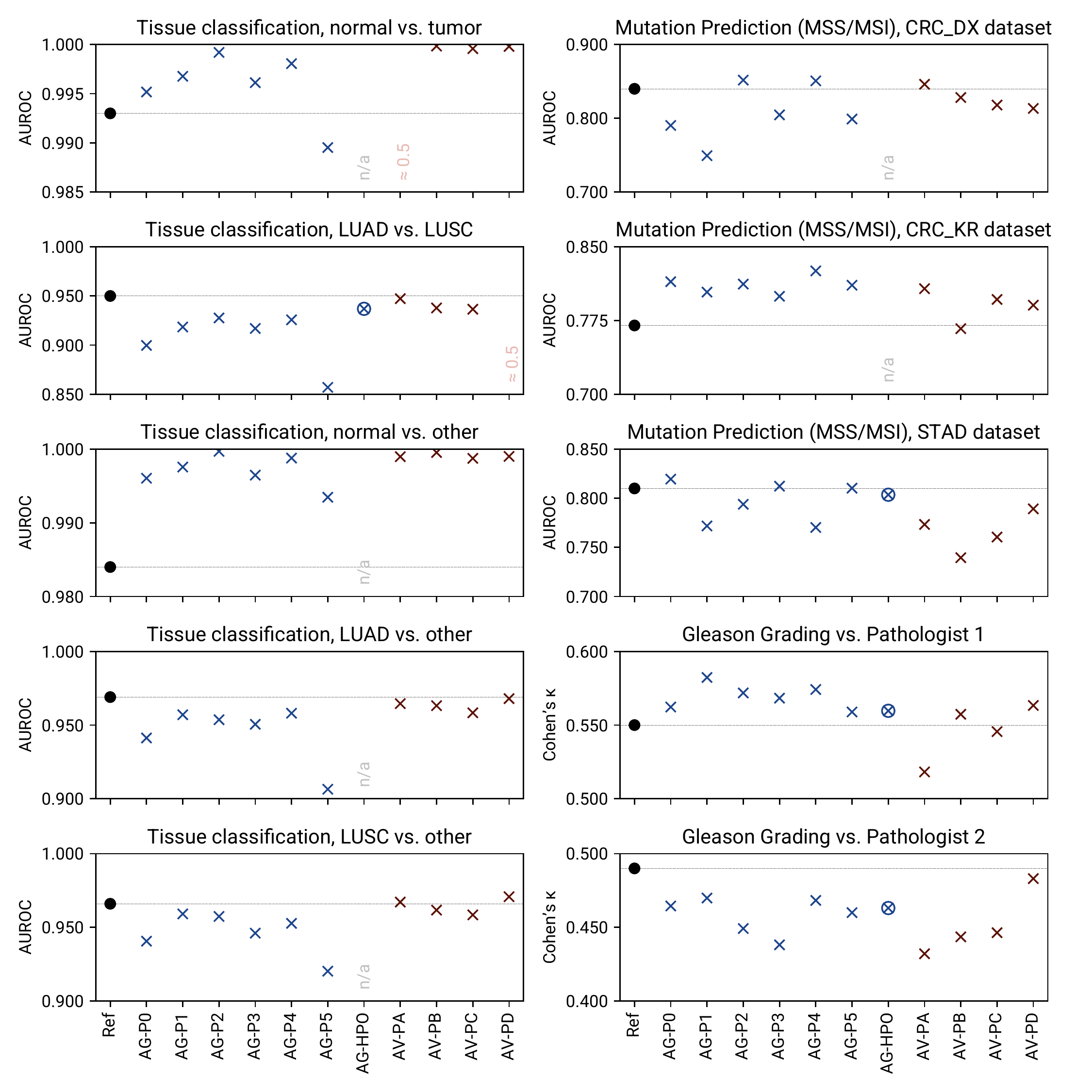}

  \caption{Overview of classifier performance. Classifier performance
    is quantified in the respective metrics for the tasks of our use
    cases, results for the AutoGluon (AG; blue crosses) and AutoML
    Vision (AV; red crosses) are compared to the reference results
    from literature (black circles). For AutoGluon, the circled cross
    indicates the results of the preset completing a limited
    hyperparameter optimization. (“n/a” indicates those tasks where the
    HPO preset was not run; “$\approx 0.5$” indicates classifiers
    returning constant classes, i.e., where the AutoML optimization
    clearly failed).}
  \label{fig:Performance}
\end{figure}

In the initial evaluation of AutoGluon, preset~2 generally generated
the best-performing classifiers, see \autoref{fig:Performance} (large
blue crosses).

For the tissue classification use case, distinguishing normal from
tumor tissue (two-class normal vs.\ tumor, three-class normal
vs.\ other) seems to be a comparably easy task, since AUROCs were
generally larger than 0.99. The AutoGluon-trained classifiers produced
AUROCs even closer to 1 than the reference results from the
literature. For the distinction between LUAD and LUSC (two-class task)
as well as for the LUAD vs.\ other and LUSC vs.\ other (three-class
task), the performance of the AutoGluon-generated classifiers was
almost on par with the literature results.

For the mutation prediction use case, the AutoGluon-generated
classifiers resulted in slightly higher AUROCs for two of the datasets
and a slightly lower AUROC for the third dataset.

For the grading use case, κ~values compared to pathologist~1 were
slightly higher and κ values compared to pathologist~2 were slightly
lower than the literature reference for the AutoGluon-generated
classifiers. As pathologist~1 created the annotations for the training
and validation data, this may indicate a slightly stronger overfitting
of the algorithm to this observer than the literature result, possibly
due to our oversampling.

Given the relatively high variability when repeating the AutoML runs
(cf.\ below), we refrained from computing potentially misleading
improvement percentages in the comparisons throughout this section.

\subsubsection{AutoML Vision}

The classifiers generated by AutoML Vision generally performed
similarly to those generated by AutoGluon and the reference results,
see \autoref{fig:Performance} (large red crosses). Unlike the
AutoGluon-generated classifiers that output floating-point numbers,
the classifiers generated by AutoML Vision output 8-bit
integer-quantized prediction probabilities. The resulting small number
of possible values causes many ties when finding the class with
highest probability. These ties were resolved in the evaluation by
using the “first” class with maximum probability, where “first”
refers to an order opaquely chosen by AutoML Vision (constant per
dataset in our observations).

Among the presets, none was generally superior to the others, in
particular not the one named “higher accuracy.” This observation
also held for the tile-based average accuracy reported after training,
even in those cases where the full test data was available for cloud
evaluation. Presets~A and~D failed to generate a meaningful classifier
in one case each, these classifiers predicted the same class for all
test tiles. For the “normal vs.\ tumor” task, this happened after
reaching the budget limit of 2 node-hours, and the classifier
constantly predicted one class with high probability (244/256). For
the “LUAD vs.\  LUSC” task, the AutoML run stopped prior to reaching
the budget limit (after 5.957 of 16 node-hours) for unknown
reasons. In this case, the classifier predicted both classes with a
probability of 0.5 for all tiles. For the tissue classification use
case, the classifiers generated by AutoML Vision resulted in AUROCs
even closer to 1 than the AutoGluon-generated classifiers for the
tasks distinguishing normal and tumor slides (two-class normal
vs.\ tumor, three-class normal vs.\ other).  For the two-class LUAD and
LUSC task as well as for the three-class LUAD vs.\ other and LUSC
vs.\ other tasks, AUROCs were slightly higher than for the
AutoGluon-generated classifiers and in the same range as the reference
results. For the mutation prediction use case, AUROCs are in the same
range as the literature results and as for the AutoGluon-generated
classifiers. For the grading use case, κ values were slightly lower
than for the AutoGluon-generated classifiers, and in the same range as
the reference results.

\subsection{Impact of hyperparameter optimization}

Our hyperparameter optimization in AutoGluon led to slightly, but not
substantially better classifier performance compared to preset~2 for
the tissue classification and mutation prediction use cases, see
\autoref{fig:Performance}. For the grading use case, the classifier
performance was slightly better than the one for preset~2 for
pathologist~2's annotations, but interestingly slightly worse for
pathologist~1's annotations.

\section{Hyperparameter optimization}

A more detailed analysis of the tile-based accuracies
(\ref{appendix:HPO}) corroborated this observation: for the tissue
classification and grading use cases
(Figures~\ref{suppFig:HPOTissueClassification} and
\ref{suppFig:HPOGrading}), validation did not substantially improve
compared to the initial trial.  While there was a substantial
improvement for the mutation prediction use case
(\autoref{suppFig:HPOMutationPrediction}); this was actually due to
the initial parametrization being different from the one in
preset~2. Among the hyperparameters optimized, learning rate had the
highest influence on classifier performance.  Generally, a larger
learning rate led to better classifier performance, so it might have
made sense to extend the search range beyond $10^{-2}$.

AutoGluon presets~1, 2, 3, and~5 also included partial hyperparameter
optimization to the extent possible within the time budget of 6 hours.
In 18 of the 60 cases, more than one trial was started. In 4 of those
18 cases, accuracy increased after the first trial, but none of these
improvements was substantial.

AutoML Vision did not report whether or not hyperparameter optimization
took place during the respective run.

\subsection{Reproducibility}

\begin{figure}[tp]
  \includegraphics[width=\textwidth]{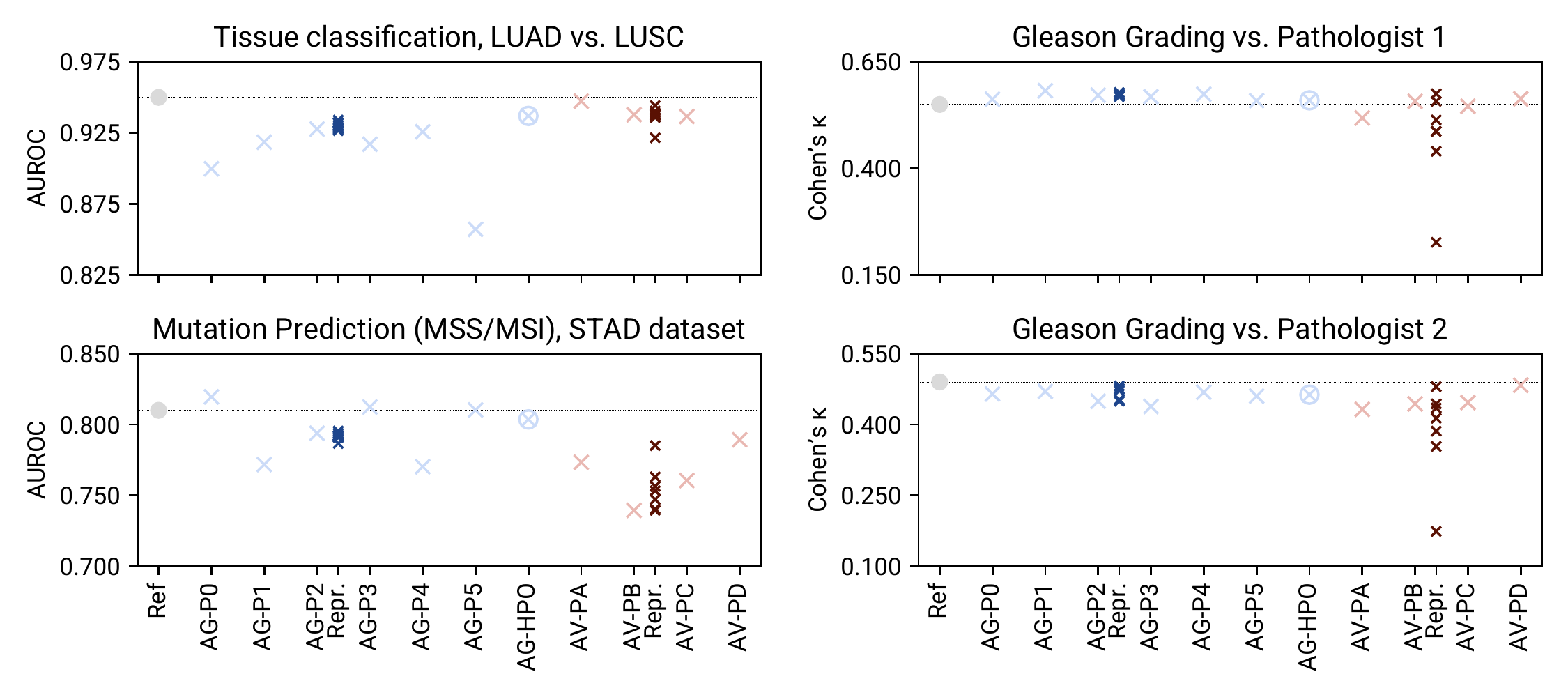}

  \caption{Variability of AutoML results. For selected tasks,
    classifier performances are shown for one AutoGluon (blue) and one
    AutoML Vision (red) preset for an additional six runs. The bright
    crosses show the results from the initial run for all presets for
    comparison.}
  \label{fig:PerformanceReproducibility}
\end{figure}

The reproducibility analysis for selected tasks (see
\autoref{fig:PerformanceReproducibility}) showed notable
variability of the performance metrics, including cases where the
first AutoML run yielded worse results than the six repetitions for
some of the tasks.

The classifiers generated by AutoML Vision were more variable than the
AutoGluon-generated ones. This variance included one particularly
obvious outlier in the comparison to both pathologists, resulting from
the same repetition. Given the small number of repetitions in this
study and the single reported reference results, we refrained from a
potentially misleading statistical significance analysis.

The variability was particularly large for the grading use case. This
was probably because the tile-based results were not aggregated per
slide, so there was no compensation for misclassifications. In fact, a
more detailed analysis of the per-tile predictions for the other use
cases also showed higher variability than the aggregated results, see
\ref{appendix:reproducibility}.

The range of values obtained from these repeated AutoML runs using the
same data cannot be directly compared to confidence intervals reported
in~\citep{CouOcaSak2018} and~\citep{KatPeaHal2019}, which were obtained
by bootstrapping the test data for fixed classifiers. We also computed
these for completeness, see \ref{appendix:bootstrappedCI}.

\section{Discussion}

\subsection{Comparison of AutoGluon and AutoML Vision}

As an open-source tool, AutoGluon allows insight into how it performs
the individual runs and permits adaptation if needed. In contrast,
AutoML Vision is a black-box tool whose internals may change without
users noticing. In both cases, the generated CNNs and their execution
environment are under the control of the user, allowing transparent
evaluation (e.g., for quality assurance purposes) of the resulting
classifiers.

AutoGluon allows setting network architectures and parametrizing the
hyperparameter optimization, whereas AutoML Vision only allows
selecting one of three presets and a run-time budget, for which a
recommendation is given based on the dataset size. This gives
AutoGluon more flexibility at the cost of higher usage complexity.

AutoGluon can be used on premises, so confidentiality and privacy of
training data can be ensured. In contrast, cloud-based tools like
AutoML Vision may be incompatible with local data protection
regulations.  On-premises tools require appropriate GPU resources and
maintenance of the corresponding systems, while cloud-based tools
incur costs for cloud data storage, data transfer, and, most importantly,
GPU time.

In the AutoML Vision evaluation for the tissue classification use
case, we observed two cases where the classifier predicted a constant
single class for the entire test set, which suggests an obviously
failed training. In contrast, the outlier in the Gleason grading
corresponds to a classifier predicting different classes (a small
proportion of which agree with the reference results). Just this
single result is insufficient to identify the corresponding classifier
as sub-optimal.

Both tools allow tile-based inference within the tool itself, i.e.,
AutoML Vision allows deploying the generated CNNs for inference in the
cloud. We decided to run this step on premises in order to be able to
better control and review the evaluation. Aggregating CNN output per
tile to an assessment per slide is a step not necessarily supported in
generic AutoML tools and, thus, needs to be implemented as a
postprocessing step in either case.

The performance of the classifiers obtained for the three selected
use-cases using the two selected AutoML tools was generally on par
with literature results. While the reference results in the literature
were obtained with network architectures and training specifically
optimized for the respective task, we obtained our classifiers using
generic presets with minimal interaction effort.

For AutoGluon, generally our preset~2 seemed to be a good choice to
obtain consistently good classifiers, both independent of the specific
use case and independent of non-deterministic effects. The limited
hyperparameter optimization evaluated here resulted in much larger
computational effort (46-fold, 6.9-fold, and 6.7-fold), but no
substantial improvement of classifier performance. For AutoML Vision,
our results do not permit a clear preset recommendation. However,
given that the resulting classifiers can sometimes be useless or
sub-optimal, it may be necessary to perform additional AutoML runs.

\subsection{Reproducibility}

The results of the reproducibility analysis illustrate two rather
fundamental effects also occurring independently of AutoML:
non-determinism in Deep Learning and the robustness of aggregated
metrics.

It is well known that training and evaluation of CNN-based classifiers
are not fully deterministic processes~\citep{PhaQiaWan2020, MorWil2020,
  NagWarSto2018, LixTal2020, LinHut2020} due to multiple effects:

\begin{itemize}
\item The system used (including but not limited to the GPU) and
  concurrent load on the system (in particular for disk
  cache/input/output and CPU) may differ.
\item In combination with a run-time budget specified in wall-clock
  time, actual computational resources may differ.
\item Seeding of pseudo-random number generators may differ (for the
  actual training, but also for pre- and postprocessing), explicit
  seeding may not be available via the AutoML tool.
\item Order of data files read from disk may vary, resulting in
  different training/validation/test splits and training order.
\item Parallelization of floating-point arithmetics on the GPU is
  inherently non-deterministic~\citep{PiaMarSan2018, MarOliPia2019}.
\end{itemize}

These effects mostly have an impact on the training due to its
multiple iterations, but may also affect inference. In particular, the
non-determinism for training runs with the same hyperparameters poses
a problem for hyperparameter optimization by making the objective
function noisy, regardless whether algorithmic or manual optimization
is used.

Our approach was to run the training under essentially the same
conditions from the perspective of a naive user of the AutoML
tools. In particular, we did not dig into the source code of AutoGluon
to work out where pseudo-random number generators seeding would be
needed, and we did not run the AutoML on a system without concurrent
jobs. For AutoML Vision, these aspects were beyond our control in the
cloud platform in any case. A more detailed investigation of the
contributions of individual effects, how they could be mitigated, and
how non-determinism interacts with hyperparameter optimization is
beyond the scope of the present study. Such an investigation could
involve separating training and inference by using single-threaded
code on the CPU for inference; seeding pseudo-random number generators
to the extent possible; using a system exclusively without concurrent
computational jobs; etc.

The larger variability at the tile level (see
\ref{appendix:reproducibility}) is actually what characterizes
reproducibility of the runs of the AutoML tools; aggregated
performance metrics may obfuscate such non-determinism. On the other
hand, this discrepancy indicates that the underlying (diagnostic)
tasks are robust with respect to misclassifications of part of the
data for separate cases. One way to mitigate or profit from this
variability is to generate multiple CNNs and use them as an
ensemble~\citep{NguBlaDaw2021, LinLee2020, KasKasWes2019,
  QumKhaSha2019, GanHuoth2021}, either at the tile level before
aggregating information per slide, or at the slide level—a detailed
investigation of which is also beyond the scope of this study.

\subsection{Limitations and outlook}

Besides the points mentioned above, the biggest limitation of this
study is the number of use cases considered and the limitation to
tasks addressable by image classification. While we selected three use
cases to cover typical tasks in computational pathology, our findings
should be confirmed or disproven in more applications, e.g., those
listed in~\citep{EchRinBri2020}.  With easy-to-use generic AutoML tools
becoming more flexible and with the development of self-configuring
approaches for medical image analysis tasks~\citep{IseJaeKoh2020,
  BauJaeIse2021}, an evaluation of these techniques for tasks in
computational pathology also seems within reach.

A second major limitation is that we only evaluated two AutoML tools.
While this is sufficient to show that AutoML tools can achieve
performance on par with reference results from literature, a
comparison with additional available tools would be interesting. This
should involve generic on-premises tools (e.g.,~\citep{web-autokeras,
  web-autopytorch, web-modelsearch}) generic cloud tools (such
as~\citep{web-AWSImageClassification, web-vertexai, web-customvision}),
and tools specific for computational pathology
(e.g.,~\citep{WanCouZha2021}).  The latter can have the advantage of
containing re-usable building blocks needed for histological images
independent of the specific diagnostic task, such as tiling of
whole-slide images and the per-slide aggregation of inference
results~\citep{JanMad2016} and corresponding specialized loss functions
or evaluation metrics, as well as targeted image modifications for
data augmentation or normalization~\citep{TelLitBan2019b}.

Within the AutoGluon experiments, we defined separate presets for
separate CNN architectures to have a clearer separation of parameters
varied during hyperparameter optimization, making sure all
architectures are tried within limited run-time. For a neural
architecture search other than trying standard architectures, a
different platform would be needed. Moreover, our hyperparameter
optimization only ran for 16 trials, using a limited set of parameters
with limited ranges in order to evaluate what is achievable within
typically available computational resources (about 11.5 days). It
would be interesting to extend both neural architecture search and
hyperparameter optimization to investigate whether substantially
larger computational effort can robustly yield higher performance than
the reference results.

Our comparison of classifier performance was based on (differences of)
data-driven metrics, AUROC and quadratically weighted Cohen’s~κ, to
allow a numerical comparison. This is purely a data science
perspective and needs to be complemented by an assessment by domain
experts (pathologists) in a suitable study with classifiers integrated
in a practical diagnostic workflow.

Further potential for domain-specific improvement lies in CNN
initialization. For image classification, CNNs are typically
pre-trained on ImageNet~\citep{DenDonSoc2009} to benefit from transfer
learning~\citep{RagZhaKle2019}.  For histological images, pre-training
on other computational pathology tasks can be beneficial in order to
exploit synergies due to specific image characteristics different from
photographs of everyday objects as contained in ImageNet. Such
approaches include autoencoders~\citep{WanYanRon2019, TelLitLaa2019c}
or training (parts of) single CNNs for multiple
tasks~\citep{TelHoeVer2020}, possibly in a federated
manner~\citep{LuxKonLip2020, SheEdwRei2020, DayRotZho2021}.

From a statistical point of view, the obtained performance values need
to be interpreted carefully. In order to have a meaningful comparison
between different presets, we used a single set of test data per task
for all presets. This might overestimate the performance for data
beyond the present datasets, a further independent test set would be
needed for a proper estimation of how the performance
generalizes. Furthermore, we largely followed the dataset splitting
strategies of the original publications in order to obtain comparable
results. This, however, results in heterogeneous approaches for the
three use cases, not always ensuring splits per patient.

The results presented here were obtained using a specific version of
AutoGluon fixed at the beginning of the in-silico experiments and the
unspecified version of AutoML Vision while our cloud AutoML
experiments were run. These results do not necessarily generalize to
later versions.

From a more general perspective, AutoML tools as investigated in this
study only address one aspect of developing solutions in computational
pathology~\citep{JanMad2016}.  Both domain and machine learning
expertise remain indispensable for properly translating between a
diagnostic task with its evaluation metrics and a machine learning
task with its associated loss and metrics. While reducing manual
efforts for optimizing CNN architectures and training procedures,
generic AutoML tools do not obviate the need for domain-specific data
curation. Detecting and addressing potential bias and imbalance in the
datasets~\citep{KomIsh2019, JohKho2019, BriMarTor2020} remains
essential. Morevoer, to ensure generalization beyond the training
data~\citep{StaEilUng2019}, suitable data
normalization~\citep{KomIsh2019, BejLitTim2016, BugSchGro2017} and
augmentation~\citep{TelLitBan2019b, KalZwoGra2011, KomIsh2019} need be
employed, also in combination with data balancing
strategies~\citep{KomIsh2019, MarSchIst2018, HuaJaf2021}.

\section{Conclusions}

The generic AutoML tools AutoGluon and AutoML Vision are capable of
generating image classifiers whose performance is on par with
literature results for selected use cases in computational
pathology. The impact of actual AutoML features is unclear; AutoGluon
relied on a generic preset for CNN architecture and training
hyperparameters. The AutoML tools thus eliminate the need to manually
search for neural architectures and optimize hyperparameters, which
can significantly reduce development effort and allow developers to
focus on harder-to-automate tasks like data curation.

\section*{CRediT authorship contribution statement}

\textbf{Lars Ole Schwen:} Conceptualization, Methodology, Software, Data Curation, Writing – Original Draft, Visualization
\textbf{Daniela Schacherer:} Software, Data Curation, Writing – Review
\textbf{Christian Geißler:} Writing – Review
\textbf{André Homeyer:} Conceptualization, Methodology, Writing – Review, Project administration, Funding acquisition

\section*{Acknowledgments}

The authors would like to thank Rieke Alpers for advice on statistics.
This study was funded by the German Federal Ministry for Economic
Affairs and Energy (BMWi) via the EMPAIA project, grant numbers
01MK20002B (LOS, DS, AH) and 01MK20002C (CG).
The sponsors had no role in the study design; in the collection,
analysis and interpretation of data, in the writing of the manuscript;
and in the decision to submit the manuscript for publication; nor did
the developers or providers of any AutoML tools.

\section*{Declaration of Competing Interest}

The authors declare that they have no known competing financial
interests or personal relationships that could have appeared to influence
the work reported in this paper.

\bibliography{literature}

\appendix

% \clearpage

\section{Code}

The code for preprocessing data, running AutoGluon to generate CNNs
(AutoGluon only), and evaluating the resulting classifiers (AutoGluon
and AutoML Vision) is available from~\cite{codeSupplement}.

\section{Details on datasets}\label{appendix:datasets}

The splitting strategies following the original publications
(described in the Methods section) resulted in the sizes of splits shown in Tables~\ref{suppTab:tileStatisticsTissueClassification}, \ref{suppTab:tileStatisticsMutationPrediction}, and
and~\ref{suppTab:tileStatisticsGrading} as well as the class balances shown in Tables~\ref{suppTab:classStatisticsTissueClassification}, \ref{suppTab:classStatisticsMutationPrediction}, and~\ref{suppTab:classStatisticsGrading}.

\begin{sidewaystable}[p]
  \centering
  \caption{Tile statistics, tissue classification use case. These
    datasets were split at the patient level (indicated by~*) For use
    with AutoML Vision, the datasets needed to be limited to 1 million
    images. As the subsequent evaluation was run separately and not
    subject to this limit, the LUAD vs.\ LUSC task used the full
    training and validation sets, the other two tasks used 820850
    training and 179050 validation tiles.}
  \label{suppTab:tileStatisticsTissueClassification}

  \begin{tabular}{lrrr@{\qquad}rrr@{\qquad}rrr}
    \toprule
                      & \multicolumn{3}{c}{Tiles}     & \multicolumn{3}{c}{Slides} & \multicolumn{3}{c}{Patients*}                             \\
    \midrule
    Task              & Training                       & Validation                     & Test                           & Training                     & Validation                  & Test                        & Training*                   & Validation*                 & Test*                       \\
    normal vs.\ tumor & \multirow{2}{*}{926335 (70\%)} & \multirow{2}{*}{202059 (15\%)} & \multirow{2}{*}{200799 (15\%)} & \multirow{2}{*}{1473 (68\%)} & \multirow{2}{*}{350 (16\%)} & \multirow{2}{*}{344 (16\%)} & \multirow{2}{*}{739 (73\%)} & \multirow{2}{*}{143 (14\%)} & \multirow{2}{*}{127 (13\%)} \\
    normal vs.\ LUAD vs.\ LUSC                                                                                                                                                                                                                                                                                 \\
    LUAD vs LUSC      & 756378 (70\%)                  & 162625 (15\%)                  & 162678 (15\%)                  & 1084 (69\%)                  & 249 (16\%)                  & 243 (15\%)                  & 719 (72\%)                  & 141 (14\%)                  & 134 (13\%)                  \\
    \bottomrule
  \end{tabular}
  \bigskip

  \caption{Tile statistics, mutation prediction use case. The test set
    was split at the patient level, whereas training and validation
    data were split at the tile level (indicated by~*)}
  \label{suppTab:tileStatisticsMutationPrediction}

  \begin{tabular}{lrrr@{\qquad}rrr@{\qquad}rrr}
    \toprule
     & \multicolumn{3}{c}{Tiles*} & \multicolumn{3}{c}{Slides} & \multicolumn{3}{c}{Patients*} \\
    \midrule
    Task (Dataset) & Training*    & Validation* & Test          & Training   & Validation & Test       & Training   & Validation & Test*      \\
    CRC\_DX        & 79397 (42\%) & 11676 (6\%) & 98904 (52\%)  & 263 (42\%) & 262 (42\%) & 101 (16\%) & 260 (42\%) & 259 (42\%) & 100 (16\%) \\
    CRC\_KR        & 51760 (38\%) & 7612 (6\%)  & 78249 (57\%)  & 504 (42\%) & 484 (41\%) & 199 (17\%) & 278 (42\%) & 271 (41\%) & 109 (17\%) \\
    STAD           & 85484 (40\%) & 12572 (6\%) & 118008 (55\%) & 199 (40\%) & 198 (40\%) & 103 (21\%) & 185 (40\%) & 184 (39\%) & 99 (21\%)  \\
    \bottomrule
  \end{tabular}
  \bigskip

  \caption{Tile statistics, Grading use case. This dataset was split
    at the TMA level (indicated by~*). Training data was subsequently
    oversampled 8-fold, effectively resulting in 201696 image
    tiles. The test data in this use case was assessed separately by
    two observers.}
  \label{suppTab:tileStatisticsGrading}

  \begin{tabular}{lrrr@{\qquad}rrr@{\qquad}rrr}
    \toprule
    & \multicolumn{3}{c}{Tiles} & \multicolumn{3}{c}{Tissue Microarrays*} & \multicolumn{3}{c}{Patients} \\
    \midrule
                & Training     & Validation & Test        & Training* & Validation* & Test*    & Training & Validation & Test \\
    TMA dataset & 25212 (79\%) & 2482 (8\%) & 4237 (13\%) & 3 (60\%)  & 1 (20\%)    & 1 (20\%) & 508      & 133        & 245  \\
    \bottomrule
  \end{tabular}
\end{sidewaystable}

\begin{table}[p]
  \centering
  \caption{Class statistics (number of tiles), tissue classification use case}\label{suppTab:classStatisticsTissueClassification}

  \begin{tabular}{lrrrr}
    \toprule
                     & \multicolumn{2}{c}{Normal vs.\ tumor} & \multicolumn{2}{c}{LUAD vs.\ LUSC}             \\
                     & Normal                               & Tumor         & LUAD          & LUSC          \\
    \midrule
    Training tiles   & 164119 (18\%)                        & 762216 (82\%) & 369683 (49\%) & 386695 (51\%) \\
    Validation tiles & 37883 (19\%)                         & 164176 (81\%) & 79691 (49\%)  & 82934 (51\%)  \\
    Test tiles       & 45510 (23\%)                         & 155289 (77\%) & 79625 (49\%)  & 83053 (51\%)  \\
    \midrule
                     & \multicolumn{3}{c}{normal vs.\ LUAD vs.\ LUSC} \\
                     & Normal                               & LUAD          & LUSC                          \\
    \midrule
    Training tiles   & 164119 (18\%)                        & 372021 (40\%) & 390195 (42\%)                 \\
    Validation tiles & 37883 (19\%)                         & 80724 (40\%)  & 83452 (41\%)                  \\
    Test tiles       & 45510 (23\%)                         & 76254 (38\%)  & 79035 (39\%)                  \\
    \bottomrule
  \end{tabular}
  \bigskip

  \caption{Class statistics (number of tiles), mutation prediction use case.}\label{suppTab:classStatisticsMutationPrediction}

  \begin{tabular}{llrr}
    \toprule
    Datset  & Split set        & MSS          & MSI          \\
    \midrule
    CRC\_DX & Training         & 39705 (50\%) & 39692 (50\%) \\
            & Validation       & 5839 (50\%)  & 5837 (50\%)  \\
            & Test             & 70569 (71\%) & 28335 (29\%) \\
    CRC\_KR & Training         & 25906 (50\%) & 25854 (50\%) \\
            & Validation       & 3780 (50\%)  & 3832 (50\%)  \\
            & Test             & 60574 (77\%) & 17675 (23\%) \\
    STAD    & Training         & 42729 (50\%) & 42755 (50\%) \\
            & Validation       & 6303 (50\%)  & 6269 (50\%)  \\
            & Test             & 90104 (76\%) & 27904 (24\%) \\
    \bottomrule
  \end{tabular}
  \bigskip

  \caption{Class statistics (number of tiles), grading use case.}\label{suppTab:classStatisticsGrading}

  \begin{tabular}{lllll}
    \toprule
                                     & 0           & 3           & 4           & 5           \\
    \midrule
    Training                         & 2076 (14\%) & 6303 (41\%) & 4541 (30\%) & 2383 (16\%) \\
    Training, augmented              & 50424       & 50424       & 50424       & 50424       \\
    augmentation/oversampling factor & 24.3-fold   & 8-fold      & 11.1-fold   & 21.2-fold   \\
    Validation                       & 666 (27\%)  & 923 (37\%)  & 573 (23\%)  & 320 (13\%)  \\
    Test (pathologist~1)             & 127 (3\%)   & 1602 (38\%) & 2121 (50\%) & 387 (9\%)   \\
    Test (pathologist~2)             & 90 (2\%)    & 861 (20\%)  & 2715 (64\%) & 571 (13\%)  \\
    \bottomrule
  \end{tabular}
\end{table}

\section{Hyperparameter optimization}\label{appendix:HPO}

\begin{figure}[p]
  \includegraphics[width=0.9\textwidth]{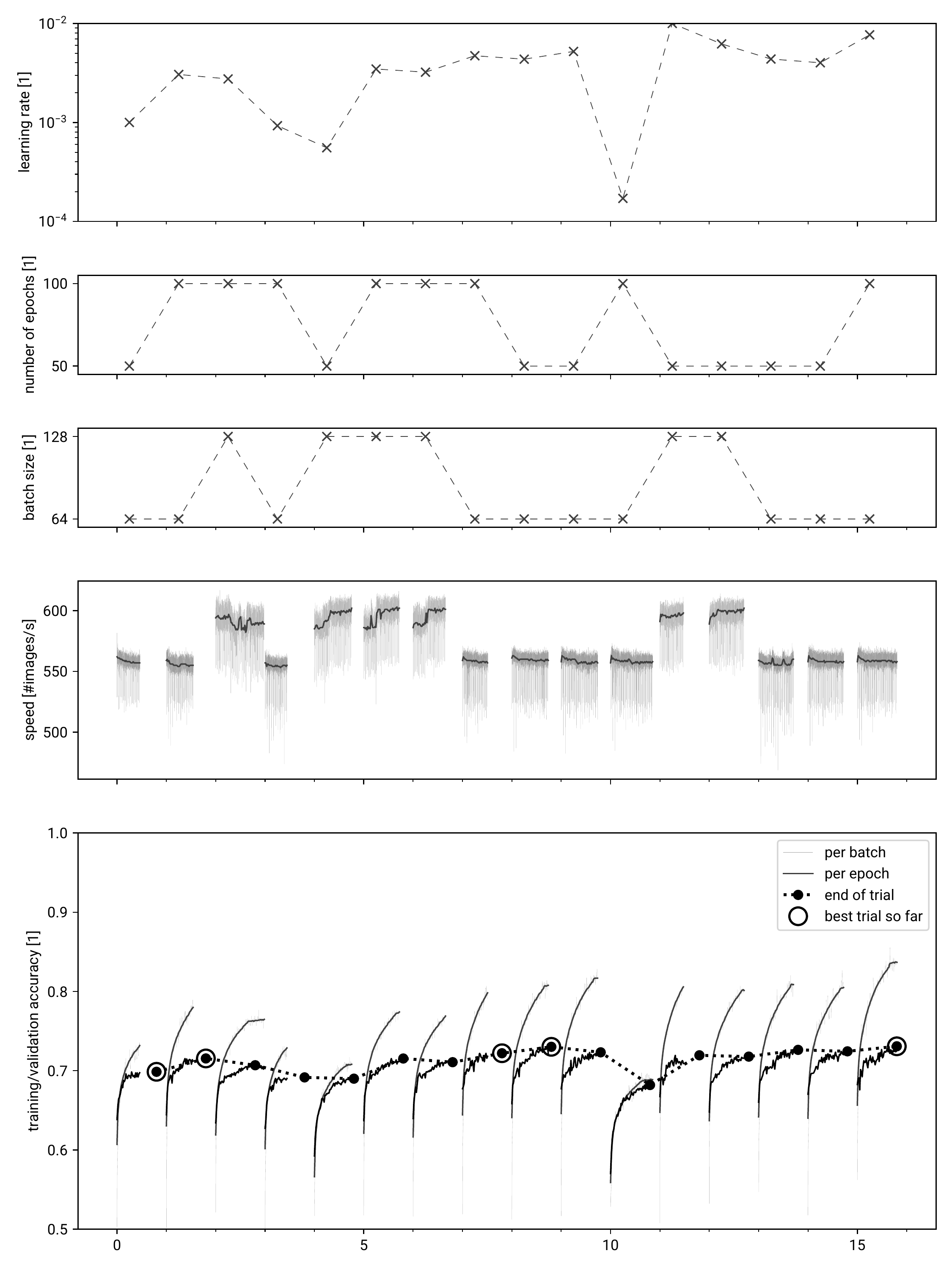}

  \caption{Training parameters and accuracies during hyperparameter
    optimization, tissue classification use case.  Thin lines:
    training data, thick lines: validation data.}
  \label{suppFig:HPOTissueClassification}
\end{figure}

\begin{figure}[p]
  \includegraphics[width=0.9\textwidth]{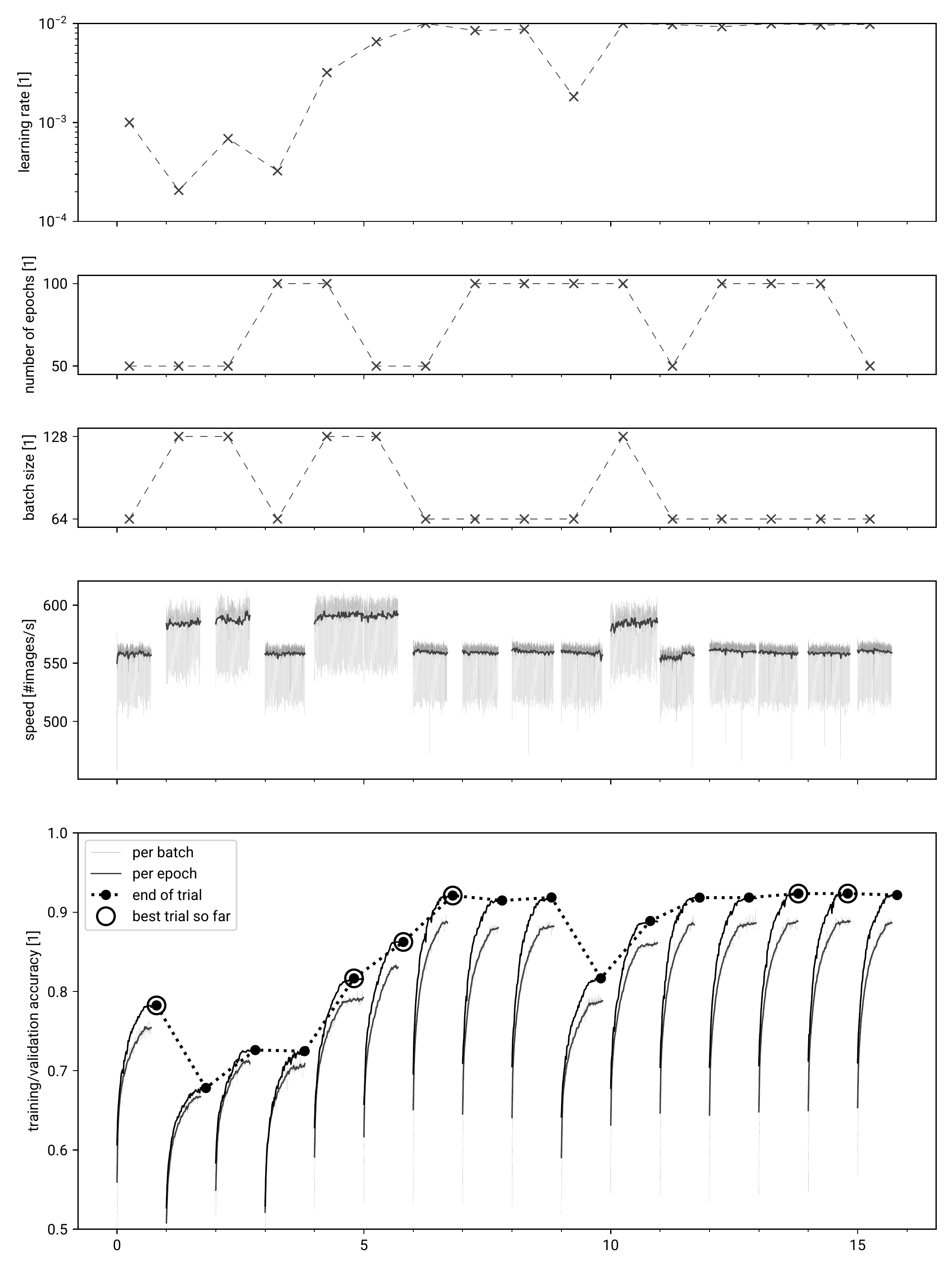}

  \caption{Training parameters and accuracies during hyperparameter
    optimization, mutation prediction use case.  Thin lines: training
    data, thick lines: validation data.}
\label{suppFig:HPOMutationPrediction}
\end{figure}

\begin{figure}[p]
  \includegraphics[width=0.9\textwidth]{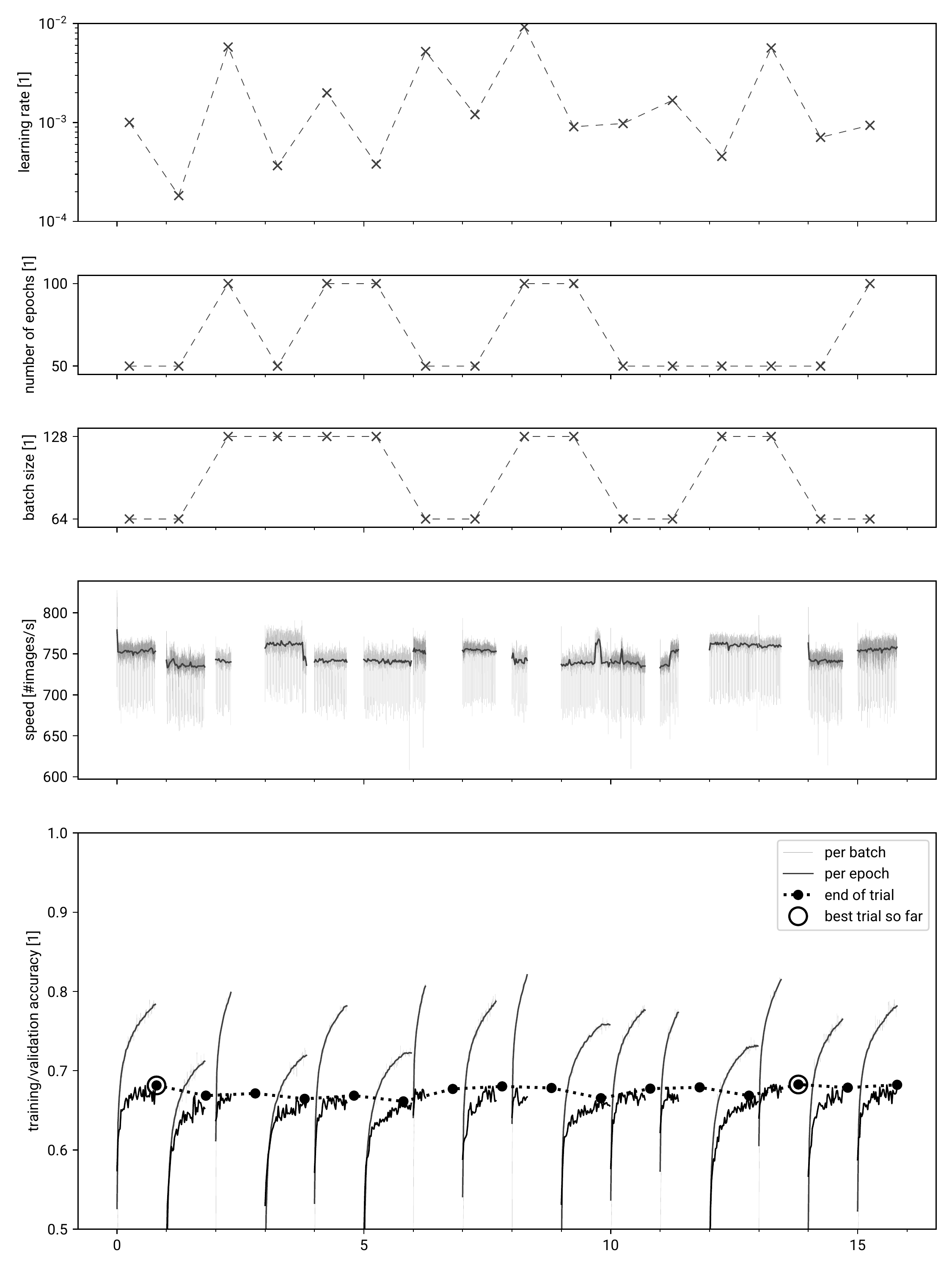}

  \caption{Training parameters and accuracies during hyperparameter
    optimization, grading use case.  Thin lines: training data, thick
    lines: validation data.}
  \label{suppFig:HPOGrading}
\end{figure}

Training parameters and accuracies during hyperparameter optimization
are shown in \autoref{suppFig:HPOTissueClassification} (tissue
classification), \autoref{suppFig:HPOMutationPrediction} (mutation
prediction), and \autoref{suppFig:HPOGrading} (grading use
case). These accuracies are tile-based accuracies and thus lower than
the metrics used for classifier performance (slide-based ROC-AUC and
tile-based quadratically weighted Cohen's~κ)

Interestingly, validation accuracies for the mutation prediction use
case (\autoref{suppFig:HPOMutationPrediction}) are consistently
higher than the training accuracies. This is the only use case where
training and validation data overlap in patients, but otherwise the
reasons for this effect are unclear.

\section{Details on reproducibility}\label{appendix:reproducibility}

We computed variabilities of metrics of intermediate steps when
aggregating tile-based classification results to slide-based
classification for a more detailed analysis: For the tile-based
classification, i.e., the immediate output of the CNNs, we computed
means, standard deviations, and coefficients of variation of the
confusion matrices over the seven repeated experiments. Moreover, we
computed means, standard deviations, and coefficients of variation of
the tile-based accuracy and AUROC in addition to the slide-based AUROC
(tissue classification use case,
Tables~\ref{suppTab:confusionMatrixTissueClassification}
and~\ref{suppTab:aggregatedPerformanceTissueClassification}; mutation
prediction use case,
Tables~\ref{suppTab:confusionMatrixMutationPrediction}
and~\ref{suppTab:aggregatedPerformanceMutationPrediction}) as well as
for the tile-based quadratically weighted Cohen's~κ (grading use case,
Tables~\ref{suppTab:confusionMatrixGrading}
and~\ref{suppTab:aggregatedPerformanceGrading}).

\begin{table}[p]
  \centering
  \caption{Variability of the confusion matrix (column: actual class, row: predicted
    class) of tile predictions, mean~±~standard deviation (coefficient
    of variation in percent) over seven experiments for the tissue
    classification use case.}
  \label{suppTab:confusionMatrixTissueClassification}

  \begin{tabular}{lll}
    \toprule
    AutoGluon     & LUAD                      & LUSC                     \\
    \midrule
    LUAD          & 61155.6 ± 1296.4 (2.1\%)  & 18469.4 ± 1296.4 (7.0\%) \\
    LUSC          & 25368.1 ± 1093.6 (4.3\%)  & 57684.9 ± 1093.6 (1.9\%) \\
    \midrule
    AutoML Vision & LUAD                      & LUSC                     \\
    \midrule
    LUAD          & 58642.0 ± 1484.4 (2.5\%)  & 20983.0 ± 1484.4 (7.1\%) \\
    LUSC          & 19541.9 ± 2122.5 (10.9\%) & 63511.1 ± 2122.5 (3.3\%) \\
    \bottomrule
  \end{tabular}
  \bigskip

  \centering
  \caption{Variability of aggregated metrics for the tissue
    classification use case.}
  \label{suppTab:aggregatedPerformanceTissueClassification}

  \begin{tabular}{lll}
    \toprule
                        & AutoGluon              & AutoML Vision          \\
    \midrule
    Accuracy tile-based & 0.731 ± 0.0060 (0.8\%) & 0.750 ± 0.0125 (1.7\%) \\
    AUROC tile-based    & 0.812 ± 0.0065 (0.8\%) & 0.830 ± 0.0133 (1.6\%) \\
    AUROC slide-based   & 0.930 ± 0.0027 (0.3\%) & 0.937 ± 0.0072 (0.8\%) \\
    \bottomrule
  \end{tabular}
\end{table}

\begin{table}[p]
  \centering
  \caption{Variability of the confusion matrix (column: actual class,
    row: predicted class) of tile predictions, mean~±~standard
    deviation (coefficient of variation in percent) over seven
    experiments for the mutation prediction use case.}
  \label{suppTab:confusionMatrixMutationPrediction}

  \begin{tabular}{lll}
    \toprule
    AutoGluon               & MSI                     & MSS                     \\
    MSI                     & 13316.9 ± 405.1 (3.0\%) & 14587.1 ± 405.1 (2.8\%) \\
    MSS                     & 17755.6 ± 769.8 (4.3\%) & 72348.4 ± 769.8 (1.1\%) \\
    \midrule
              AutoML Vision & MSI                     & MSS                     \\
    \midrule
    MSI                     & 7884.0 ± 359.1 (4.6\%)  & 20020.0 ± 359.1 (1.8\%) \\
    MSS                     & 10026.1 ± 618.0 (6.2\%) & 80077.9 ± 618.0 (0.8\%) \\
    \bottomrule
  \end{tabular}
  \bigskip

  \centering
  \caption{Variability of aggregated metrics for the mutation
    prediction use case.}
  \label{suppTab:aggregatedPerformanceMutationPrediction}

  \begin{tabular}{lll}
    \toprule
                        & AutoGluon              & AutoML Vision          \\
    \midrule
    Accuracy tile-based & 0.726 ± 0.0034 (0.5\%) & 0.747 ± 0.0037 (0.5\%) \\
    ROC-AUC tile-based  & 0.709 ± 0.0017 (0.2\%) & 0.651 ± 0.0121 (1.9\%) \\
    ROC-AUC slide-based & 0.792 ± 0.0027 (0.3\%) & 0.755 ± 0.0158 (2.1\%) \\
    \bottomrule
  \end{tabular}
\end{table}

\begin{table}[p]
  \centering
  \caption{Variability of the confusion matrix (column: actual class,
    row: predicted class) of tile predictions, mean~±~standard
    deviation (coefficient of variation in percent) over seven
    experiments for the grading use case.}
  \label{suppTab:confusionMatrixGrading}

  \begin{tabular}{lllll}
    \toprule
    \rlap{AutoGluon} & \multicolumn{4}{c}{Test vs.\ P1}                                               \\
      & 0                    & 3                     & 4                     & 5                     \\
    \midrule
    0 & 67.7 ± 5.9 (8.7\%)   & 47.7 ± 7.1 (14.8\%)   & 9.6 ± 2.2 (22.9\%)    & 2.0 ± 1.1 (53.5\%)    \\
    3 & 27.1 ± 5.1 (18.7\%)  & 1095.9 ± 47.2 (4.3\%) & 459.6 ± 50.1 (10.9\%) & 19.4 ± 6.2 (31.9\%)   \\
    4 & 28.6 ± 10.1 (35.4\%) & 441.1 ± 58.0 (13.2\%) & 1443.7 ± 73.0 (5.1\%) & 207.6 ± 34.9 (16.8\%) \\
    5 & 6.4 ± 2.1 (32.0\%)   & 18.1 ± 3.9 (21.7\%)   & 165.9 ± 18.2 (10.9\%) & 196.6 ± 17.7 (9.0\%)  \\
    \midrule
      & \multicolumn{4}{c}{Test vs.\ P2}                                                              \\
      & 0                    & 3                     & 4                     & 5                     \\
    \midrule
    0 & 57.3 ± 3.0 (5.2\%)   & 24.9 ± 4.2 (17.0\%)   & 6.7 ± 1.5 (22.1\%)    & 1.1 ± 0.8 (72.9\%)    \\
    3 & 27.1 ± 4.6 (17.6\%)  & 647.0 ± 19.1 (3.0\%)  & 181.7 ± 20.4 (11.2\%) & 5.1 ± 2.9 (57.3\%)    \\
    4 & 35.3 ± 11.7 (33.1\%) & 876.3 ± 79.0 (9.0\%)  & 1580.1 ± 94.1 (6.0\%) & 223.3 ± 37.2 (16.7\%) \\
    5 & 10.1 ± 3.6 (35.5\%)  & 54.7 ± 11.6 (21.1\%)  & 310.1 ± 21.4 (6.9\%)  & 196.0 ± 18.5 (9.4\%)  \\
    \midrule
    \rlap{AutoML Vision} & \multicolumn{4}{c}{Test vs.\ P1}                                           \\
      & 0                    & 3                     & 4                     & 5                     \\
    \midrule
    0 & 185.0±77.2 (41.7\%)  & 52.0±47.0 (90.5\%)    & 137.0±54.1 (39.5\%)   & 13.0±6.4 (49.5\%)     \\
    3 & 0.3±0.5 (170.8\%)    & 107.0±10.7 (10.0\%)   & 2.1±2.3 (109.2\%)     & 17.6±10.0 (57.1\%)    \\
    4 & 231.0±141.6 (61.3\%) & 185.3±106.8 (57.6\%)  & 1279.6±218.6 (17.1\%) & 425.1±237.8 (55.9\%)  \\
    5 & 22.6±16.8 (74.5\%)   & 186.1±75.2 (40.4\%)   & 367.4±154.6 (42.1\%)  & 1025.9±168.6 (16.4\%) \\
    \midrule
      & \multicolumn{4}{c}{Test vs.\ P2}                                                              \\
      & 0                    & 3                     & 4                     & 5                     \\
    \midrule
    0 & 197.4±83.0 (42.0\%)  & 57.4±47.3 (82.4\%)    & 277.6±63.1 (22.7\%)   & 38.6±22.7 (58.9\%)    \\
    3 & 0.3±0.5 (170.8\%)    & 79.0±5.1 (6.5\%)      & 1.9±2.0 (109.6\%)     & 8.9±4.3 (49.0\%)      \\
    4 & 236.1±150.3 (63.6\%) & 252.4±136.0 (53.9\%)  & 1369.7±292.8 (21.4\%) & 856.7±311.8 (36.4\%)  \\
    5 & 5.0±5.2 (103.9\%)    & 141.6±48.5 (34.3\%)   & 137.0±64.1 (46.8\%)   & 577.4±81.3 (14.1\%)   \\
    \bottomrule
  \end{tabular}
  \bigskip

  \caption{Variability of the aggregated metric, quadratically
    weighted Cohen's~κ, for the grading use case.}
  \label{suppTab:aggregatedPerformanceGrading}

  \begin{tabular}{lll@{\qquad}ll}
    \toprule
      & \multicolumn{2}{c}{AutoGluon}                   & \multicolumn{2}{c}{AutoML Vision}               \\
      & Test vs.\ P1            & Test vs.\ P2            & Test vs.\ P1            & Test vs.\ P2            \\
    \midrule
    κ & 0.573 ± 0.0034 (0.6\%) & 0.465 ± 0.0129 (2.8\%) & 0.469 ± 0.117 (24.8\%) & 0.384 ± 0.101 (26.3\%) \\
    \bottomrule
  \end{tabular}
\end{table}

\section{Bootstrapped confidence intervals}\label{appendix:bootstrappedCI}

\begin{figure}[p]
  \includegraphics[width=\textwidth]{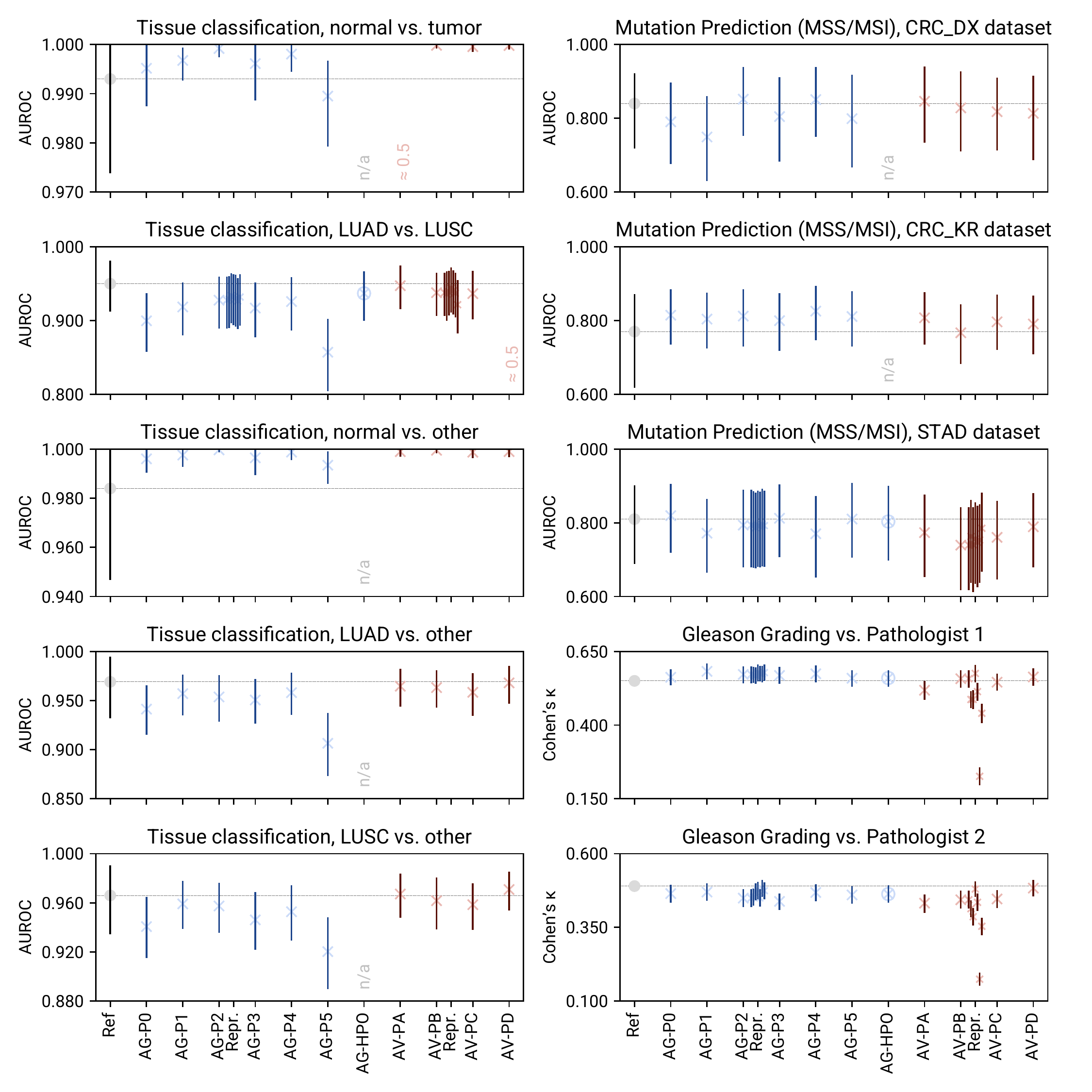}

  \caption{95\%~confidence intervals obtained by 2000-fold
    bootstrapping over the per-slide predictions (tissue
    classification and mutation prediction use cases) and per-tile
    predictions (grading use case)}
  \label{suppFig:PerformanceWithCI}
\end{figure}

In~\citep{CouOcaSak2018}, the authors computed 95\% confidence
intervals estimating how well the trained CNNs would generalize beyond
the available data by 1000-fold bootstrapping of the per-slide
predictions, resulting in ranges of (0.974, 1.000) for normal
vs.\ tumor, (0.913, 0.980) for LUSC vs.\ LUAD, and (0.947,
1.000)/(0.933, 0.994)/(0.935, 0.990) for normal vs.\  rest/LUAD
vs.\ rest/LUSC vs.\ rest. In~\citep{KatPeaHal2019}, the authors used the
same approach as in~\citep{CouOcaSak2018} to compute 95\% confidence
intervals by 2000-fold bootstrapping over the test data, resulting in
(0.72, 0.92) for CRC-DX, (0.62, 0.87) for CRC-KR, and (0.69, 0.90) for
STAD. No such confidence intervals are reported for the tile
classification in~\citep{ArvFriMor2018}.  We used the same approach
with 2000-fold bootstrapping to obtain confidence intervals for all
three use cases, shown in \autoref{suppFig:PerformanceWithCI}.

\end{document}